# Normal-mode splitting of four-level atom-cavity system under collective strong coupling

Zheng Tan[1,2], Liyong Wang[1,2,3], Min Liu[1,2], Yifu Zhu[2,4], Jin Wang[1,2] and Mingsheng Zhan[1,2,*]

[1] *State Key Laboratory of Magnetic Resonance and Atomic and Molecular Physics, Wuhan Institute of Physics and Mathematics, Chinese Academy of Sciences, Wuhan 430071, China*

[2] *Center for Cold Atom Physics, Chinese Academy of Sciences, Wuhan 430071, China*

[3] *School of Physical Science, University of Chinese Academy of Sciences, Beijing 100049, China*

[4] *Department of Physics, Florida International University, Miami, Florida 33199, USA*

∗ *mszhan@wipm.ac.cn*

We investigate the transmission spectrum of an optical cavity coupled with four-level atoms. Multiple normal-mode splitting peaks of the strongly coupled atom-cavity system are obtained as single cavity mode couples three separated atomic transitions simultaneously. We employ a confocal optical cavity and an ensemble of cold $^{85}$Rb atoms for cavity quantum electrodynamics. We observe four normal-mode splitting peaks of the cavity transmission in the strong coupling regime due to collective enhancement of the atom-cavity coupling strength by ensemble of cold atoms. The experimental observations are consistent with the theoretical analysis. The multiple normal-mode excitation in the strongly coupled multi-level atom-cavity system may lead to practical application for realizing multi-channel all-optical devices.

# I. INTRODUCTION

As a signature of strong coupling cavity quantum electrodynamics (CQED), normal-mode splitting appears in a two-level atom resonantly coupled to a single cavity mode, which is also called vacuum Rabi splitting in the low-intensity limit. Normal-mode splitting in an ensemble of two-level atoms has been demonstrated in various systems, such as thermal atoms [1-3], cold atoms [4-8] and solid-state materials [9, 10]. For a two-level atomic system, the normal-mode splitting is characterized by the atom-cavity coupling strength $g=\mu\sqrt{\omega_c/2\hbar\varepsilon_0 V}$, where $\mu$ is the atom dipole matrix element, $\omega_c$ the resonant frequency of the cavity and $V$ the cavity mode volume. The atom-cavity coupling strength is increased to $g\sqrt{N}$ when cavity modes are collectively coupled with N atoms (N is the number of atoms interacting with the cavity modes). Thus, large collective coupling strength in a coupled atom-cavity system can be achieved by confining a large amount of atoms in cavity modes, which is desirable to quantum metrology applications [11, 12]. Recently, the study of collective coupling of atom and cavity has been extended to from single mode to multimode CQED which includes multiple longitudinal or transverse cavity modes. When the collective atom-cavity coupling strength $g\sqrt{N}$ is comparable to or larger than the cavity free spectral range $\Delta_{FSR}$, the situation called superstrong coupling regime is achieved. Normal-mode splitting peaks of different cavity longitudinal modes in this regime were observed [13-16]. Moreover, mode-resolved cavity transmission spectrum for atoms coupled with several cavity transverse modes in a nearly confocal optical cavity has also been experimentally demonstrated [17].

Meanwhile, the study has also been extended for atoms from two-levels to multi-levels. There has been considerable interest for systems composing of N multi-level atoms cooperatively interacting with single mode laser field, because they are crucial platforms for both fundamental study and practical applications. When three-level atoms are resonantly coupled to a single cavity mode, the transmission spectrum of the cavity exhibits three-peaked structures, which has been widely studied in the cavity electromagnetically induced transparency (EIT) system [18-23]. Recently, coherent effect in multi-level atomic systems induced by a single laser field becomes attractive topic, e.g. in a laser-driven four-level atomic system which consists of three closely spaced upper levels and a common lower level, atomic coherence can be created via spontaneously generated coherence (SGC) when different decay pathways are coupled with the same vacuum modes [24-27]. However, to the best of our knowledge, similar effect that n-level atoms (n>3) coherently coupled with a single cavity mode is not been investigated in the atom-cavity system. It has been shown that photon entanglement can be generated via the cascade transition processes in the CQED system with ensemble of multi-level atoms [28, 29]. So, it is desirable to extend the study of the multi-atom CQED to a new regime in which multiple atomic transitions coupled with a single cavity mode are included.

In this work, we present the experimental observation and theoretical analysis of the transmission spectrum of an optical cavity strongly coupled with ensemble of four-level atoms. In comparison with the case of cavity coupled two-level atomic system, the excitation spectrum of the atom-cavity system changes dramatically and multiple normal-mode splitting peaks appear as the result of increased coupling channels between cavity modes and multi-level atoms. By coupling with D2 line transition of cold $^{85}$Rb atoms, we show that the excitation spectrum of the CQED system presents four normal-mode splitting peaks as the cavity detuning is varied to match different atomic transitions. Such cavity mode coupling simultaneously with multiple atomic transitions may offer new ways for quantum state manipulation and be used for constructing all-optical devices with multi-channel output.

This work is organized as follows. In Sec. II we present a semi-classical theoretical model of a single cavity mode coupled with four-level atoms and interpret the generation of the multiple normal-mode peaks in the cavity transmission spectrum. In Sec. III we describe the experimental setup and procedure. In Sec. IV, we report the experimental results with a composite atom-cavity system containing cold $^{85}$Rb atoms and an optical cavity. The potential applications of the results are also discussed. Finally, we draw a conclusion in Sec. V.

## II. THEORETICAL ANALYSIS

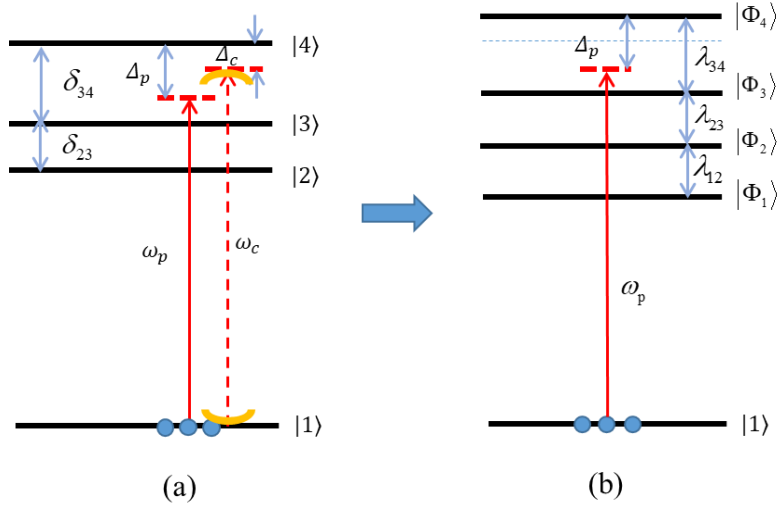

Fig.1. The schematic diagrams of a cavity mode coupled with four-level atoms. (a) The cavity mode simultaneously couples the atomic transitions $|1\rangle \rightarrow |2\rangle, |1\rangle \rightarrow |3\rangle$ and $|1\rangle \rightarrow |4\rangle$. (b) The energy level diagram of the coupled atom-cavity system containing ground state $|1\rangle$ and four normal modes $|\Phi_1\rangle, |\Phi_2\rangle, |\Phi_3\rangle$ and $|\Phi_4\rangle$ in the dressed state representation.

Figure 1 shows the schematic diagram of the coupled atom-cavity system, N four-level atoms are confined in a single mode cavity. A weak probe laser field with amplitude $a_{in}$ is coupled from one side of the cavity and excites the atom-cavity system. The cavity mode couples with three atomic transitions $|1\rangle - |2\rangle, |1\rangle - |3\rangle$ and $|1\rangle - |4\rangle$ simultaneously, and the cavity frequency detuning is defined as $\Delta_c = \nu_c - \nu_{41}$, where $\nu_c$ is the cavity resonant frequency and $\nu_{41}$ is the frequency separation between states

$|1\rangle$ and $|4\rangle$. The cavity probe laser has the frequency of $\nu_p$, which is detuned from the atomic transition $|1\rangle-|4\rangle$ by $\Delta_p = \nu_p - \nu_{41}$. The frequency separation between states $|4\rangle$ and $|3\rangle$, $|3\rangle$ and $|2\rangle$ are $\delta_{34} = \omega_4 - \omega_3$ and $\delta_{23} = \omega_3 - \omega_2$, respectively.

The Hamiltonian for the coupled atom-cavity system is:

$$H = \hbar \sum_{m=1}^{N}\sum_{i=1}^{4} \omega_i \sigma_{ii}^{(m)} + \hbar \hat{a}^+ \hat{a} + \hbar(\sum_{m=1}^{N}\sum_{i=1}^{4} g_i \hat{a} \sigma_{i1}^{(m)} + i\hat{a}^+ \sqrt{2\kappa_l}\, a_{in} + H.C.). \tag{1}$$

Where $\hat{\sigma}_{ii}^{(m)}$ is the atomic population operator of the $m$-th atom in state $|i\rangle$; $\hat{\sigma}_{i1}^{(m)}$ is the off-diagonal atomic operators of the $m$-th atom which represents atomic coherence between states $|1\rangle$ and $|i\rangle$. $\hat{a}\,(\hat{a}^+)$ is the annihilation (creation) operator of the cavity photons, $a_{in}$ is the amplitude of the probe laser field coupled into the cavity, $g_i = \mu_{1i}\sqrt{\omega_c/2\hbar\varepsilon_0 V}$ (i=2-4) is the atom-cavity coupling strength. $\kappa_j = T_j/\tau$ (j=r or l) is the loss rate of the cavity on the mirror $j$ ($T_j$ is the mirror transmission and $\tau$ is the photon round trip time inside the cavity). The equations of motion for the coupled atom-cavity system are given by $\frac{d\hat{\rho}}{dt} = \frac{1}{i\hbar}[\hat{H},\hat{\rho}] + \hat{L}\hat{\rho}$ [30], where $\hat{L}$ is the quantum super operator and represents the system decay. For $N$ atoms confined in the cavity mode, we assume that the coupling strength is equal and the laser field is uniform, thus $\hat{\sigma}_{lm}^{(m)} = \hat{\sigma}_{lm}$. Under the rotating wave approximation, we obtain equations of motion for $\hat{\sigma}_{lm}$ and the intra-cavity field $\hat{a}$ as follows [31]:

$$\dot{\sigma}_{11} = \sum_{i=2}^{N}[\gamma_i \sigma_{ii} + i(g_i a \sigma_{i1} - g_i a^+ \sigma_{1i})] \tag{2}$$

$$\dot{\sigma}_{22} = \gamma_2 \sigma_{22} - i(g_2 a \sigma_{21} - g_2^* a^+ \sigma_{12}) \tag{3}$$

$$\dot{\sigma}_{33} = \gamma_3 \sigma_{33} - i(g_3 a \sigma_{31} - g_3^* a^+ \sigma_{13}) \tag{4}$$

$$\dot{\sigma}_{44} = \gamma_4 \sigma_{44} - i(g_4 a \sigma_{41} - g_4^* a^+ \sigma_{14}) \tag{5}$$

$$\dot{\sigma}_{12} = -[\frac{\gamma_2}{2} + i(\Delta_p + \delta_{23} + \delta_{34})]\sigma_{12} + ig_2 a^+(\sigma_{22} - \sigma_{11}) + ig_3 a^+ \sigma_{32} + ig_4 a^+ \sigma_{42} \tag{6}$$

$$\dot{\sigma}_{13} = -[\frac{\gamma_3}{2} + i(\Delta_p + \delta_{34})]\sigma_{13} + ig_3 a^+(\sigma_{33} - \sigma_{11}) + ig_2 a^+ \sigma_{32} + ig_4 a^+ \sigma_{43} \tag{7}$$

$$\dot{\sigma}_{14} = -[\frac{\gamma_4}{2} + i\Delta_p]\sigma_{14} + ig_4 a^+(\sigma_{44} - \sigma_{11}) + ig_2 a^+ \sigma_{24} + ig_3 a^+ \sigma_{34} \tag{8}$$

$$\dot{\sigma}_{23} = -[\frac{\gamma_2 + \gamma_3}{2} - i\delta_{23}]\sigma_{23} + ig_2^* a^+ \sigma_{13} - ig_3 a \sigma_{21} \tag{9}$$

$$\dot{\sigma}_{34} = -[\frac{\gamma_3 + \gamma_4}{2} - i\delta_{34}]\sigma_{34} + ig_3^* a^+ \sigma_{14} - ig_4 a \sigma_{31} \tag{10}$$

$$\dot{\sigma}_{24} = -[\frac{\gamma_2 + \gamma_4}{2} - i\delta_{23} - i\delta_{34}]\sigma_{24} + ig_2^* a^+ \sigma_{14} - ig_4 a \sigma_{21} \tag{11}$$

$$\dot{a} = -[(\kappa_r + \kappa_l)/2 + (\Delta_c - \Delta_p)]a + ig_2 N\sigma_{21} + ig_3 N\sigma_{31} + ig_4 N\sigma_{41} + \sqrt{\kappa/\tau} a_{in} \tag{12}$$

where $\gamma_i = \Gamma$ is the decay rate of the excited state $|i\rangle$. We assume the cavity is symmetric, thus $\kappa_r = \kappa_l = \kappa$. The quantum fluctuation terms are negligible, and $\hat{\sigma}_{i1}$ are treated as c numbers. In the weak excitation regime, the atoms nearly populate in the ground state $|1\rangle$, thus $\sigma_{11} = 1$ and $\sigma_{ii} = 0$ (i=2-4). The steady-state solution of the intra-cavity field is given by:

$$a = \frac{\sqrt{\kappa/\tau} a_{in}}{\kappa + i(\Delta_c - \Delta_p) - i\chi} \tag{13}$$

so we obtain the cavity transmitted probe field:

$$a_t = \sqrt{\kappa\tau} a = \frac{\kappa a_{in}}{\kappa + i(\Delta_c - \Delta_p) - i\chi}. \tag{14}$$

where the atomic susceptibility $\chi$ is given by:

$$\chi=\frac{ig_2^2N}{\frac{\gamma_2}{2}-i(\Delta_p+\delta_{34}+\delta_{23})}+\frac{ig_3^2N}{\frac{\gamma_3}{2}-i(\Delta_p+\delta_{34})}+\frac{ig_3^2N}{\frac{\gamma_4}{2}-i\Delta_p} \qquad (15)$$

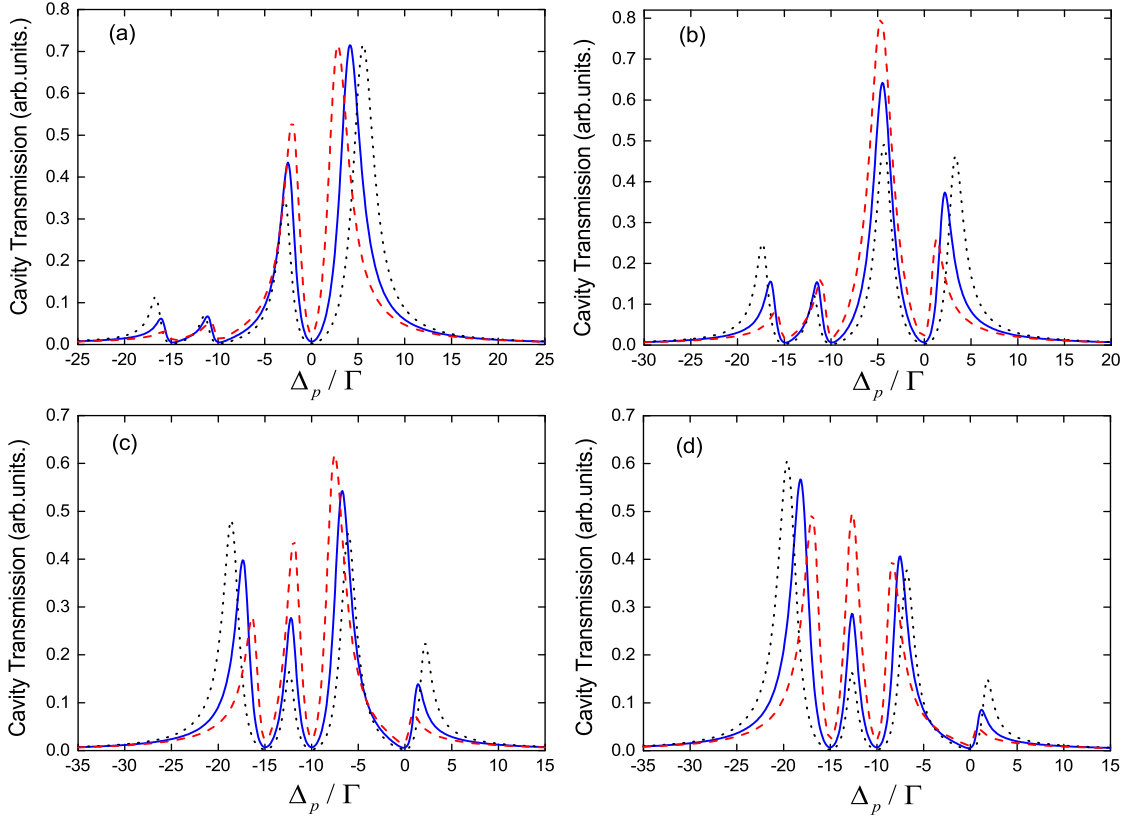

Fig. 2. Cavity transmission versus probe laser detuning $\Delta_p/\Gamma$. The black dotted line: $g\sqrt{N}$=4.3 $\Gamma$; The blue line: $g\sqrt{N}$=3.3 $\Gamma$; The red dashed line: $g\sqrt{N}$=2.3 $\Gamma$. (a) $\Delta_c$=0; (b) $\Delta_c$=-5 $\Gamma$; (c) $\Delta_c$=-10 $\Gamma$; (d) $\Delta_c$=-12.5 $\Gamma$. The other parameters are $\kappa$=2$\Gamma$, $\delta_{23}$=5 $\Gamma$, $\delta_{34}$=10 $\Gamma$.

We take $g_2=g_3=g_4=g$ in the following analysis. Owing to the increased coupling channels of cavity mode and multi-level atoms, the cavity transmission spectrum changes dramatically compared to the case of two-level atoms confined in cavity modes. Fig. 2 plots the calculated results of the transmission intensity versus the probe laser detuning $\Delta_p/\Gamma$ for different values of the cavity detuning $\Delta_c$. The spectral peaks of the cavity transmission real the energy structure of the normal modes of the coupled atom-cavity system. The spectral line shape and peak position varies with the collective coupling strength $g\sqrt{N}$ and cavity detuning $\Delta_c$. Fig. 2(a) corresponds to the cavity frequency is tuned to be resonant with the atomic transition $|1\rangle \rightarrow |4\rangle$, that is the cavity detuning $\Delta_c = 0$. For the strong coupling regime of the CQED system, the transmitted peak of the empty cavity splits into four peaks representing the four normal modes excitation of the atom-cavity system. The peaks located in the vicinity of the cavity frequency have much higher intensity because of the relatively strong coupling strength, while the amplitude of peaks far away from the cavity frequency become weaker. Fig. 2(b) shows the transmission spectrum when the cavity frequency is tuned to the middle of the $|3\rangle$ and $|4\rangle$ states. The amplitude of peak corresponding to the normal mode $|\Phi_3\rangle$ is higher than that of other peaks (corresponding to normal modes $|\Phi_1\rangle$, $|\Phi_2\rangle$ and $|\Phi_4\rangle$ respectively) for the reason that the cavity frequency is deviated from the atomic resonance. Fig. 2(c) shows the cavity frequency is tuned to be resonant with the atomic transition $|1\rangle \rightarrow |3\rangle$. In Fig. 2(d), the cavity detuning is set at the middle of the $|2\rangle$ and $|3\rangle$ states, the coupling strength with the $|1\rangle \rightarrow |4\rangle$ transition becomes much weaker, so the associated peak intensity also become weak. Moreover, the peak intensity changes greatly between four normal-mode eigenstates with different collective coupling strength $g\sqrt{N}$, the dependence of the cavity transmission spectrum on $g\sqrt{N}$ is indicated in Fig. 2.

In order to reveal the origin of the cavity transmission spectral features and physics of single cavity mode coupled with four-level atomic system, we describe the collective coupling of the CQED system by using the Dicke-type atomic and photonic basis states. The product states of the coupled atom-cavity system are: $|\Phi_1\rangle = \frac{1}{\sqrt{N}} \sum_{j=1}^{N} |1,...,1^j,...1\rangle|1_p\rangle$, (all atoms in the ground state $|1\rangle$ and one photon in the cavity mode); $|\Phi_2\rangle = \frac{1}{\sqrt{N}} \sum_{j=1}^{N} |1,...,2^j,...1\rangle|0_p\rangle$ ( one atom is in the excited state $|2\rangle$ and no photon in the cavity mode); $|\Phi_3\rangle = \frac{1}{\sqrt{N}} \sum_{j=1}^{N} |1,...,3^j,...1\rangle|0_p\rangle$ (one atom is in the excited state $|3\rangle$ and no photon in the cavity mode); $|\Phi_4\rangle = \frac{1}{\sqrt{N}} \sum_{j=1}^{N} |1,...,4^j,...1\rangle|0_p\rangle$ ( one atom is in the excited state $|4\rangle$ and no photon in the cavity mode); while $|1_p\rangle$ ($|0_p\rangle$) represents one (zero) photon state of the intra-cavity field. Solving the Schrödinger equation $H|\psi\rangle_\lambda = \lambda_i |\psi\rangle_\lambda$ with the interaction Hamiltonian and four basis states mentioned above and taking $\delta_{23}=2\delta, \delta_{34}=4\delta$, we have:

$$\begin{pmatrix} 0 & 0 & 0 & g_4^*\sqrt{N} \\ 0 & 4\delta & 0 & g_3^*\sqrt{N} \\ 0 & 0 & 6\delta & g_2^*\sqrt{N} \\ g_4\sqrt{N} & g_3\sqrt{N} & g_2\sqrt{N} & -\Delta_c \end{pmatrix} |\psi\rangle_\lambda = |\lambda\rangle_i |\psi\rangle_\lambda$$

(16)

Assume $g_2=g_3=g_4=g$ and solve the eq. (15), we derive:

$$\lambda^4+(\Delta_c+10\delta)\lambda^3+(24\delta^2-3g^2N-10\delta\Delta_c)\lambda^2+(20\delta g^2N+24\delta^2\Delta_c)+24g^2N\delta^2=0$$
(17)

As an example, we discuss the case that $g\sqrt{N}=10\Gamma$ and $\delta=2.5\Gamma$. Figure 3 plots the calculated four eigenvalues $\lambda_1$, $\lambda_2$, $\lambda_3$ and $\lambda_4$ versus the cavity detuning $\Delta_c/\Gamma$. Four energy eigenvalues correspond to the four normal modes of the coupled atom-cavity system, which indicate the resonant positions of atomic and photonic excitation. When the cavity frequency is near the atomic resonance, the collective excitation of CQED system consists of the superposition of the photonic product state $|\Phi_1\rangle$ and atomic product state $|\Phi_2\rangle$, $|\Phi_3\rangle$ and $|\Phi_4\rangle$. While the cavity frequency is tuned far away from the sates $|\Phi_2\rangle$, $|\Phi_3\rangle$ and $|\Phi_4\rangle$, the photonic product state and three atomic product state become decoupled with the increasing of the cavity detuning. The four eigenstates of the CQED system distribute asymmetrically relative to three atomic excited states due to different frequency separations $\delta_{23}$ and $\delta_{34}$, which is also the origin of the asymmetrical excitation spectra shown in Fig. 2.

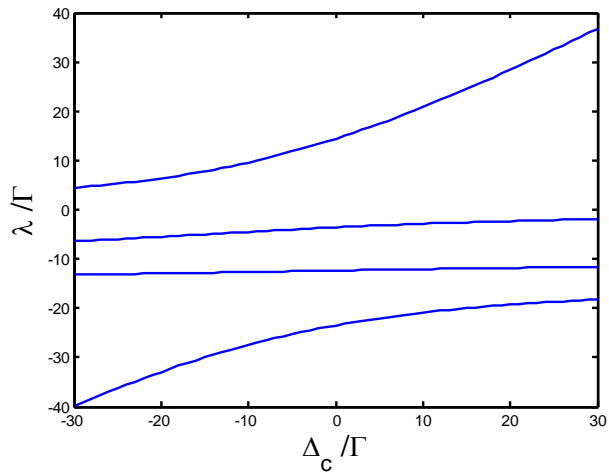

Fig. 3. The calculated eigenvalues $\lambda_1/\Gamma$, $\lambda_2/\Gamma$, $\lambda_3/\Gamma$ and $\lambda_4/\Gamma$ versus the cavity detuning $\Delta_c/\Gamma$. The other parameters are $g\sqrt{N}=10\Gamma, \kappa=2\Gamma, \delta_{23}=5\Gamma, \delta_{34}=10\Gamma$.

## III. EXPERIMENTAL SETUPS

The experimental setup is shown in Fig.4. A confocal optical cavity is used, which composes of two mirrors with same radius of curvature of 5 cm and a separation of 5 cm assembled into a 10-port stainless-steel vacuum chamber. One of the mirrors mounts on a piezoelectric transducer, so that the resonant frequency of the cavity can be tuned to match the atomic hyperfine transitions. The empty cavity finesse is measured to be ~ 300 and the measured cavity linewidth is $\kappa \approx 10 MHz$. $^{85}$Rb atoms are firstly cooled and trapped at the center of the vacuum chamber. A tapered amplifier (New focus TA-7613) with a maximum output power of ~1000 mW provides the trapping lasers which consist of three perpendicular retro-reflected beams. Another extended cavity diode laser (Toptica 100) with output power of ~ 15 mW serves as the repump laser. A getter source is used to produce the natural Rb atoms which can be controlled by electricity.

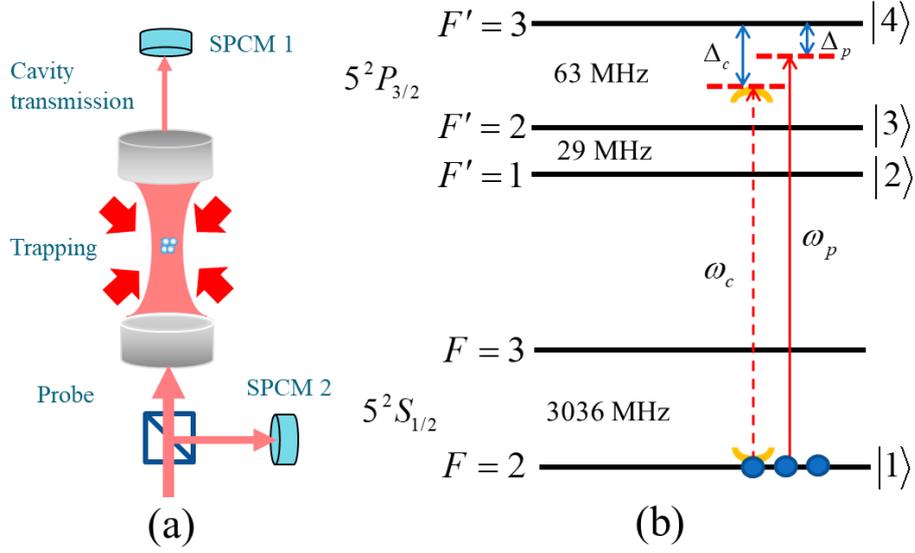

Fig.4. Experimental details. (a) Schematic of the experimental setup of the single cavity mode coupled four-level atoms. (b) The energy-level scheme for the cavity mode coupled $^{85}$Rb atoms.

The diameter of the cold atomic cloud is ~2 mm, the density of trapped $^{85}$Rb atoms can be varied by changing the laser frequency detuning from the recycling transition ($F=3 \to F'=4$). The position of the cold atomic cloud in the cavity can be finely adjusted to coincide with the center of the optical cavity by using a pair of movable anti-Helmholtz coils. A third ECDL is adopted as the probe laser, which is attenuated and coupled into the cavity by a mode-matching lens with the focal length of 15 cm. After passing through the cold atomic cloud, the transmitted light is collected using another lens with the same focal length and then coupled into a multimode fiber, the output light at the end of the fiber is detected by the single photon counting module (Excelitas SPCMAQR-15-FC).

The experiment is arranged running sequentially with a repetition rate of 10 Hz. All laser pulses are generated by acousto-optic modulators

(AOMs), trigged by the time sequence described below. For each period of 100 ms, ~ 98.8 ms are used for the cooling and trapping $^{85}$Rb atoms, during which the trapping and the repump lasers are turned on while the cavity probe laser is kept off. The repump laser is always off during the data collection procedure which lasts ~ 1.1 ms, while the trapping laser has a delayed turned off time of ~ 0.1 ms. The current to the anti-Helmholtz coils of the MOT is always kept on in the whole procedure. Then, the cavity probe laser is turned on and its frequency is scanned across the $^{85}$Rb $F=2 \rightarrow F'=1,2,3$ transition. The probe laser transmitted through the cavity is then recorded versus the probe frequency detuning.

## IV. EXPERIMENTAL RESULTS AND DISCUSSIONS

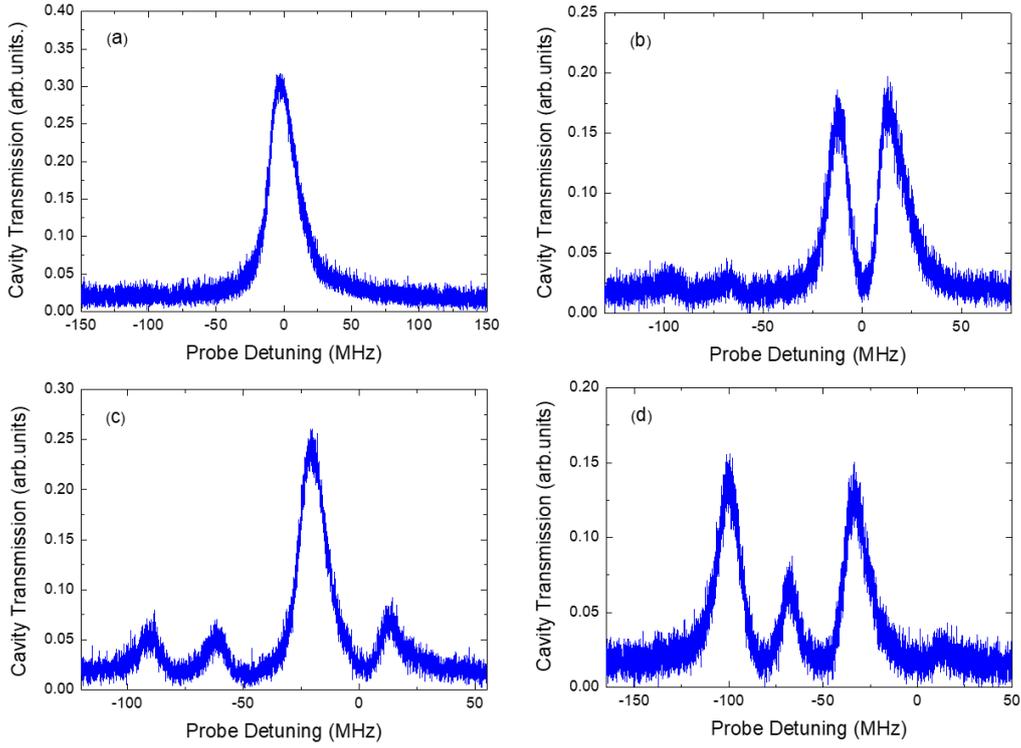

Fig.5. The transmission spectra of cavity with different cavity-atom detuning $\Delta c/\Gamma$. (a) without atom in the cavity; (b-c): with atoms in the cavity. (b) $\Delta c=0$; (c) $\Delta c= -31.7$ MHz; (d) $\Delta c= -78.1$ MHz. The other parameters are $\kappa=10$ MHz, $\delta_{23}=31.7$ MHz, $\delta_{34}=60.3$ MHz.

The cavity transmission spectra versus the probe frequency detuning are shown in Fig. 5. With varying the cavity detuning, the intensity of the transmitted peaks change accordingly. For a confocal optical cavity, each longitudinal peak contains series of transverse electromagnetic $TEM_{nm}$ modes, all those modes contribute to the normal-mode splitting of the CQED system. The atom-cavity coupling strength is considered as the average of all the contribution of $TEM_{nm}$ modes. By realizing good mode matching of the incident laser and cavity, most of the high-order modes can be eliminated, thus the $TEM_{00}$ mode dominates in the cavity modes. In this case, the cavity can be considered working at the single mode. Fig. 5 (a) shows the empty cavity transmitted peak when there is no atom in the cavity, which exhibits a Lorentzian line profile. As collective coupling strength $g\sqrt{N}$ increasing, the cavity mode can couple more than one atomic transitions simultaneously, generating multiple normal-mode slitting peaks as shown in Fig.5 (b)-(d). When the cavity frequency is tuned to resonant with the $F=2 \rightarrow F'=3$ transition, the cavity transmission spectrum is depicted in Fig.5 (b). Compared with the two-level atomic system, there appears two additional normal-mode peaks in the lower frequency due to increased coupling channels of cavity mode with the $F=2 \rightarrow F'=1$ and $F=2 \rightarrow F'=2$ transition respectively. Fig. 5 (c) shows the normal-mode peaks when the cavity frequency is detuned from the $F=2 \rightarrow F'=3$ transition with $\Delta_c \approx -31.7 MHz$ (at the middle of the $F'=2$ and $F'=3$ states), in which case the cavity mode can couple three atomic transitions: $F=2 \rightarrow F'=1$, $F=2 \rightarrow F'=2$ and $F=2 \rightarrow F'=3$ simultaneously. Fig.5 (d) presents the case when the cavity frequency is detuned from the $F=2 \rightarrow F'=3$ transition with $\Delta_c \approx -71.8 MHz$ (at the middle of the $F'=2$ and $F'=3$ state), the cavity frequency is far away from the $F'=3$ state and the separation of $F'=1$ and $F'=2$ state is smaller than $g\sqrt{N}$, so the coupling with $F'=1$ and $F'=2$ state is much stronger than with $F'=3$ state. The separation of the normal-mode peaks depends on $g\sqrt{N}$ and cavity detuning $\Delta_c$, which is described in the section II.

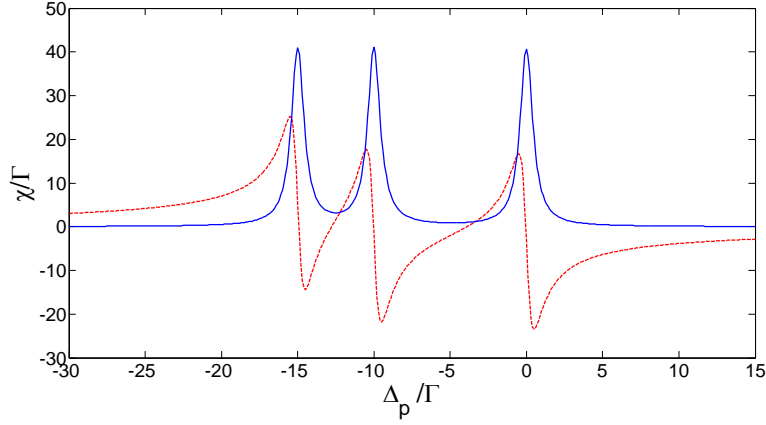

Fig. 6. The calculated atomic susceptibility $\chi/\Gamma$ versus the probe detuning $\Delta_p/\Gamma$. The blue line: the imaginary part of the susceptibility, $\chi''/\Gamma$; The red dashed line: the real part of the susceptibility, $\chi'/\Gamma$; The other parameters are $g\sqrt{N}=4.5\Gamma$, $\Delta_c=0$, $\kappa=2\Gamma$, $\delta_{23}=5\Gamma$, $\delta_{34}=10\Gamma$.

The calculated atomic susceptibility of coupled four-level atom-cavity system is shown in Fig. 6. It can be seen that the real part of the susceptibility $\chi'$, which is responsible for the dispersion of the atomic medium has a steep slope near the atomic resonance due to the narrow spectral linewidth, leading to sharp refractive index changes of atomic medium. The coupled atom-cavity system can be further manipulated with atomic coherence and quantum interference via confining the EIT-type medium in the cavity mode. For example, by coherently controlling the excitation of the coupled atom-cavity system with an additional control laser field, it may offer practical applications in realizing more efficient all-optical devices such as multi-channel optical switch and phase modulator for optical and quantum information processing.

The multiple normal-mode excitation (atom-cavity polariton) in the coupled multi-level atom-cavity system can also be applied to cavity cooling, which can cool atoms or molecules with an optical cavity by two-photon momentum transfer in the coherent scattering process [32-36]. In the strong coupling regime of two-level atom-cavity system, the enhanced cooling force [37] and the enlarged velocity capture range [38] of cavity cooling can be obtained at the atom-cavity polariton resonance. The experimental results can be used to further increase the velocity capture range of cavity cooling, owing to more velocity groups of moving atoms are involved resonating with multiple atom-cavity polariton states ($|\Phi_1\rangle$, $|\Phi_2\rangle$, $|\Phi_3\rangle$ and $|\Phi_4\rangle$). Combining with additional cooling beams, three-dimensional cavity cooling can be achieved by using one optical cavity [33], while maintaining a large velocity capture range.

Finally, it is noted that the experimental results look similar to the SGC effect in the free space, which also can give rise to narrow and multi-peak structure of spectral lines in the resonance florescence spectrum of multi-level atoms excited by the single mode laser field [24-27]. However, the existence of SGC requires very closely spaced upper levels and non-orthogonal dipole matrix elements of the atomic system which is difficult to be met in real atoms. The multi-peak structure generated in the studied CQED system can be controlled by the collective atom-cavity coupling strength, so the experimental realization is much less rigorous than that of SGC.

## V. CONCLUSION

In conclusion, we have investigated the collective excitation spectrum of a composite system consisting of an optical cavity and N four-level atoms. Multiple normal-mode splitting of the atom-cavity system is obtained due to the strong coupling of the single cavity mode with three separated atomic transitions simultaneously. The multiple normal-mode excitation of the CQED system can be viewed as the superposition of the atomic excitation and photonic excitation by adopting Dicke-type basis states, the peak positions of the excitation spectrum reveal the frequencies of the atom-cavity polariton resonance. In addition, by changing the cavity detuning $\Delta_c$ and collective coupling strength $g\sqrt{N}$, the intensities of different coupling channels can be controlled separately. We have carried out the experiment with cold $^{85}$Rb atoms confined in a confocal optical cavity, four normal-mode splitting peaks of the coupled system are observed. The experimental results are in agreement with the theoretical analysis. By coherent controlling of the multiple excitations of atom-cavity polariton, the presented results may further be applied to optical multiplexing and photon routing.

ACKNOWLEDGMENT

This work is supported by the Strategic Priority Research Program of the Chinese Academy of Sciences (CAS) (XDB21010100) and the National Natural Science Foundation of China (NSFC) (11404375); Y.Z. acknowledges support from the National Science Foundation (NSF) (1205565).